# Lifestyle Tradeoffs and the Decline of Societal Well-being: An Agent-based Model

Chris Thron

Department of Sciences and Mathematics, Texas A&M University – Central Texas, 1001 Leadership Place, Killeen TX USA 76549

thron@tamuct.edu

**Abstract.**

This paper presents a semi-quantitative mathematical model of the changes over time in the statistical distribution of well-being of individuals in a society. The model predicts that when individuals overvalue the more socially conspicuous aspects of well-being in their lifestyle choices, then the average well-being of the overall population may experience continuous decline. In addition to tradeoff cost and overvaluation, we identify statistical variation in individuals' well-being and turnover within the population as key factors driving negative trends. We investigate the influence of the effects of heterogeneity in the population, as well as economic and/or technological progress.

## Introduction

Several studies show that increases in a society's material and technological prosperity do not necessarily bring corresponding increases in well-being or happiness. During the decade from 1995 to 2005, the mean per-capita income in mainland China rose by 150 percent, while studies report the mean level of self-reported well-being (SWB) dropped significantly during the same period (Burkholder n.d.; Wong 2006). Japan from 1958 to 1987 saw a 400 percent increase in real income, with no significant increase in average self-reported happiness level (Easterlin 1995). Similarly, the U.S. experienced strong economic growth from 1946-1990, while some indicators showed a decrease in happiness (Lane 1999). Diener and Oishi (2000) reported that among 15 industrialized nations over an average of 16 years, only four showed significant increases in SWB (two actually showed significant *decreases*) during a period where average annual economic growth was 2.4 percent. The much-discussed "Easterlin paradox" asserts that across a variety of countries there is no significant increase in SWB with increasing GDP. Recent studies have argued for (Easterlin et. al. 2010) and against (Deaton 2008; Stevenson and Wolfers 2008) Easterlin's assertion. In any event, it seems clear that whether or not significant increases in SWB do occur, they are often not commensurate with the enormous gains in material prosperity resulting from economic and technological development.

Various explanations for this phenomenon have been proposed. Some authors attribute such results to rising expectation levels which increase as rapidly as real income (Graham 2009).Such perpetual striving for attainments above what have been achieved is referred to as a "hedonic treadmill" (Brickman and Campbell 1971). Some psychologists have theorized that each individual possesses a stable level of SWB (referred to as the "set point") around which the individual's happiness fluctuates (Fujita and Diener 2005). Others cite "relative deprivation", and contend that those that get wealthier still find themselves increasingly worse off relative to those they consider to be their peers (Brockmann et. al. 2009).

In this paper we propose an empirically-based mathematical model of individuals' decision-making within a society that gives a plausible account of observed non-improvements in SWB, as well as negative social trends (such as rising crime levels and family instability) that often accompany strong economic development. We verify the model with agent-based simulations. According to the model, the cumulative effect of individuals' free choices may under some circumstances produce decreases in the actual well-being of the population as a whole.

Before we present the model, some caveats are in order. SWB is only one possible measure of individuals' well-being. Some research indicates that there is a significant difference between SWB and other measures of "actual" well-being (Kahneman and Kruger 2006). Furthermore, well-being is multidimensional, and difficult to characterize in a single index (Ryff 1989). We will not attempt to define actual well-being precisely—but we do assert that the factors we discuss should be relevant to any measure of well-being that includes both material and non-material aspects.

We also emphasize that although our model is mathematical, it is not intended to be quantitatively precise. Rather, our aim is to show how socioeconomic factors may interact synergistically to produce various trends in a society's well-being, and to suggest possible modification strategies to reverse pernicious trends and/or enhance positive ones. As such, we are interested in directions and comparative sizes of effects induced by variations in salient factors, rather than exact quantification of well-being levels under different circumstances.

The model in this paper builds on the agent-based model introduced in (Thron 2014). In this paper we simplify the model, enlarge its scope, and give a much more thorough characterization of its behavior.

## Model specification

### Assumptions

The model presented in this paper is based on the following assumptions, which are supported by the sociological literature:

a. Each decision-maker in the society makes lifestyle choices in such a way as to improve the anticipated well-being of the decision-maker and those that (s)he is responsible for. This amounts to a presumption of "rational choice", which is frequently held in economic and sociological modeling (Lindenberg, 1992).

b. The different factors that contribute to a decision-maker's well-being are more or less "conspicuous", in that there are varying degrees to which they are apparent to his/her social connections (including friends, extended family, coworkers, neighbors, and so on), and enhance his/her prestige and/or social status.[1] Besides the purchase of luxury items (Vigneron & Johnson, 1999), conspicuous factors may include participation in clubs (Phillips 1994, Charles 1993), churches, (Goode 1966), children's sports leagues (Siegenthaler & Gonzalez 1997), and so on. Examples of less-conspicuous factors include aspects of "social capital" such as personal friendships and family or neighbor relationships (Coleman 1988), "serious leisure" such as hobbies (Stebbins 2011), sense of purpose, peace of mind, and personal spirituality. Also to be included among less-conspicuous factors are commonly-shared environmental conditions such as cleanliness, beauty, peacefulness, and safety (Hardin 1968), which in developing countries are often neglected in favor of economic development (Ascher & Healey 1990).

c. Decision-makers tend to overestimate the effect of conspicuous factors on well-being as compared to inconspicuous factors. Many sociological researchers have remarked on the influence of style and prestige on economic decisions, apart from considerations of functional utility (Rae 1834, Veblen 1899, Leibenstein 1950, Di Giovinazzo and Naimzada 2015). Advertising often appeals to consumers' sense of prestige and luxury (Dubois & Czellar 2002). Lifestyle decisions are often strongly influenced by the perceived impression that they will make on others (Erving 1959).

d. At any given time, available lifestyle choices reflect current conspicuous norms. Employers offer competitive salaries and benefits, based on current job market conditions (Milkovich, Newman & Gerhart 2014); in the economic literature, the "hedonic model" explains housing prices in terms of external housing characteristics (Sirmans, Macpherson & Zietz 2005); automobile pricing is also competitive and hedonic (Berry, Levinsohn, & Pakes, 1995); and the same can be reasonably supposed for other substantial consumer purchases.

e. Typically, lifestyle choices involve tradeoffs between various factors, some of which are more conspicuous than others. As a result, more conspicuously advantageous options tend to also have more inconspicuous disadvantages. For example, decisions to spend money to increase ostensible standard of living also tend to increase financial pressures and inner stress (Prawitz et. al. 2006). Career advancement may come at the expense of family commitment (Blair-Loy 2009). In addition, lifestyle options that are conspicuously desirable for one individual may have negative spillover to others: for instance, the decision to buy a luxury vehicle that pollutes more and is more obstructive in traffic.

---

[1] Some references (Liebenstein 1950, Vigneron & Johnson 1999) distinguish between "bandwagon", "snob", and "Veblen" factors associated with social conformity, social distinctiveness, and social status respectively. We will refer to all of these as "conspicuous" factors, since they all have to do with the opinion of others rather than utility to the individual.

**Mathematical specification of the basic model**

In view of the above assumptions, we constructed an agent-based model of well-being in a society. The society is characterized as consisting of $N$ "agents", where each "agent" is an entity that makes decisions which affect its own well-being. The notion of agent is flexible enough to represent various decision-making scenarios, such as a head of household deciding for his/her own family, or a couple making joint decisions, as well as an individual making personal decisions that affect only him/herself.

In this section we provide the mathematical specifications for a basic version of the model, and later consider possible variations. The next section explains how these specifications exemplify the assumptions stated in the previous section.

1. At each discrete time t=0,1,2,3,…, (we suppose that time is measured in years) the *relative well-being* of each agent $n$ is denoted by $W_t(n)$, and is determined by a *conspicuous well-being index* $C_t(n)$ and an *inconspicuous well-being index* $I_t(n)$ as follows:

$$W_t(n) \equiv C_t(n) + I_t(n).$$

2. The initial index values $(C_0(n), I_0(n))$ are chosen randomly and independently for each agent $n$ according to a bivariate normal distribution with mean (0,0). The co-variance ellipse (which indicates the region of maximum probability density) for this bivariate normal distribution is as shown in Figure 1, where $k>0$ is the *tradeoff rate*, and $\sigma$ and $\phi\sigma$ are the major and minor axis standard deviations, respectively, where $0 \leq \phi \leq 1$. Note the major axis of the covariance ellipse lies along the line $y=-kx$ in the conspicuous-inconspicuous plane.

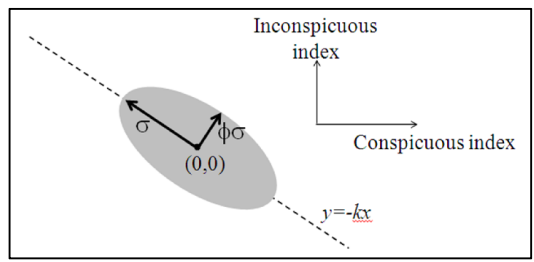

**Figure 1 Covariance ellipse for model**

3. The *anticipated relative well-being* of agent $n$ at time $t$ is denoted by $A_t(n)$, and is given by

$$A_t(n) \equiv (1+q)C_t(n) + I_t(n),$$

where $q$ is called the *overvaluation factor*.

4. At each time step $t=0, 1, 2, 3\ldots$ a fraction $f$ of the agents in the population encounter the possibility of a significant lifestyle change. For each of these agents, the new lifestyle is associated with conspicuous and inconspicuous well-being indices which are chosen according to a bivariate normal distribution with the same covariance as (2) above, but with means given by:

Mean (conspicuous, inconspicuous) indices for new lifestyle choices at time $t$ = $(m_C, -km_C),$

where $k$ is the tradeoff rate and $m_C$ denotes the current conspicuous index averaged over all agents:

$$m_C \equiv [\, C_t(1) + C_t(2) + \ldots + C_t(N)\, ]/N.$$

5. Each agent $n$ that encounters a new lifestyle possibility at time $t$ will definitely accept if the new choice improves its anticipated well-being. In other words, if $(c,i)$ represents the new lifestyle choice available to agent $n$ at time $t$, then the agent will certainly accept this choice if $(1+q)c + i > A_t(n)$. Otherwise, the agent accepts the new choice with probability $p$, where $p$ is called the *turnover probability* and is the same for all agents. In summary, the conspicuous and inconspicuous well-being indices of agent $n$ are updated as follows:

$$(1+q)c + i > A_t(n) \Rightarrow (C_{t+1}(n), I_{t+1}(n)) = (c,i),$$

$$(1+q)c + i \leq A_t(n) \Rightarrow (C_{t+1}(n), I_{t+1}(n)) = \begin{cases} (c,i) & \text{with probability } p, \\ (C_t(n), I_t(n)) & \text{with probability } 1-p \end{cases},$$

where $(c,i)$ is the new lifestyle possibility offered to agent $n$ at time $t$.

**Explanation of mathematical specifications**

The mathematical specifications (1–5) are related to the empirically-based assumptions (a–e) as follows.

The conspicuous and inconspicuous indices defined in (1) reflect respectively the sum total of conspicuous factors and inconspicuous factors that contribute to relative well-being, as described in assumption (b). $W_t(n)$ may be interpreted as the relative well-being of agent $n$ compared to the average well-being of all agents at time $t=0$, since according to (2) the average value of $W_0(n)$ is equal to 0.

Each agent's anticipated well-being $A_t$ (defined in (2)) also takes both conspicuous and inconspicuous factors into account. But since $q>0$, agents place more weight on conspicuous factors than is warranted by their actual contribution to well-being. This is in accordance with assumption (c).

The covariance matrix for $(C_0(n), I_0(n))$ in (2) produces a negative correlation between conspicuous and inconspicuous well-being for new lifestyle choices, since $k>0$. This reflects the tradeoff between conspicuous and inconspicuous factors articulated in assumption (e). Note that if $\phi=1$, then $C_0(n)$ and $I_0(n)$ are uncorrelated; and when $\phi=0$, the correlation coefficient is -1. The larger the value of $k$, the greater the tendency of conspicuous gains to produce a negative impact on inconspicuous well-being. The value of $k$ will depend on the particular circumstances of the society in question. It is also possible that $k$ may depend on time, or on conspicuous well-being itself. These possibilities are discussed later in the paper.

According to (4), the available lifestyle choices $(c,i)$ at each time $t$ have mean conspicuous index equal to the current average conspicuous index. This reflects assumption (d) that available lifestyle choices reflect current conspicuous norms. On the other hand, the mean inconspicuous index decreases with current average conspicuous index, reflecting the tradeoff between conspicuous and inconspicuous described in basic assumption (e). The covariance matrix for $(c,i)$ is the same as that for $(C_0(n), I_0(n))$, and for similar reasons.

According to (5), each agent will definitely accept a lifestyle choice that improves its *anticipated* well-being, in accordance with assumption (a). On the other hand, lifestyle choices that worsen an agent's anticipated well-being are still accepted with turnover probability $p$. This aspect of the model accounts for a variety of practical effects. First, agents may be forced to accept a less-desirable option due to personal misfortune. Also, the turnover probability accounts for the fact that agents are continually entering and leaving the population, so that experienced agents that have already "moved up the ladder" are replaced with neophyte agents that are in the process of establishing themselves will not be as discriminating in their choices.

**Limitations**

Admittedly, the assumptions of the model are vastly oversimplified. In particular, all agents in the model are faced with the same distribution of lifestyle choices. In this respect, the model more accurately reflects the situation of a socioeconomically-homogeneous subpopulation within a larger population. Besides this, the model fails to capture many of the complications involved in the socioeconomic evolution of a real-world population. As stated in the Introduction, our goal is not to provide a comprehensive model, but rather to characterize the synergy between certain factors and to investigate the effects to be expected when these factors' conditions are changed.

## Behavior of the basic model

**Preliminary characterization of model parameters**

The model has seven parameters. Of these, the number of agents $N$ has little effect on the evolution of the distribution of well-being, as long as $N$ is sufficiently large. The well-being variance $\sigma^2$ determines the numerical well-being scale—we may in fact consider the values of well-being to be measured in units of $\sigma$. The fraction undergoing lifestyle change $f$ only affects the time scale over which changes take place. Since neither $N$, $\sigma^2$, nor $f$ affect the qualitative behavior of the distribution of well-being, we fix these three parameters at the following values: $N$=10,000, $\sigma^2$=1, and $f$=0.2, and focus on the behavioral effects of the four remaining parameters: overvaluation factor $q$, tradeoff rate $k$, covariance factor $\phi$, and turnover probability $p$. The choice of $f$=0.2 was not verified by any empirical studies, but it seems plausible that roughly 20% of a population makes major lifestyle changes during a given year. Accordingly, in the following simulations the time scale is denoted as "years". By this time measure, it appears that some of the simulations below are run for very long time periods (up to 200 years).This was done so that the distributions for the different parameter-value scenarios would be clearly separated in the figures. Since the trends are constant, our conclusions are valid for shorter time intervals as well.

**Behavior with zero covariance factor**

We first look at the effects of $k$, $q$, and $p$ in the simple case where $\phi$=0. In this case, it is particularly easy to understand the model dynamics.

Figure 2 shows the results of a set of simulations in which $q$=0.75, and the tradeoff rate $k$ takes four different values corresponding to $k$<1, 1<$k$<1+$q$, $k$=1+$q$, and $k$>1+$q$. For each different $k$ value the rate of change in well-being is constant, thus yielding a straight line—but somewhat surprisingly, the slopes of these lines first decrease, then increase as $k$ is increased. Remarkably, the largest rate of increase corresponds to the *largest* tradeoff rate ($k$=2.25), which corresponds to the *fastest* decrease of inconspicuous well-being with increasing conspicuous well-being. On the other hand, relative well-being actually decreases when 1<$k$<1+$q$. When $k$=1+$q$, there is a very slight increase in relative well-being over time. When a turnover probability of $p$=0.1 is introduced, the slopes of the well-being lines are only slightly affected. Note that a positive turnover probability actually *benefits* the agents when 1< $k$ <1+$q$.

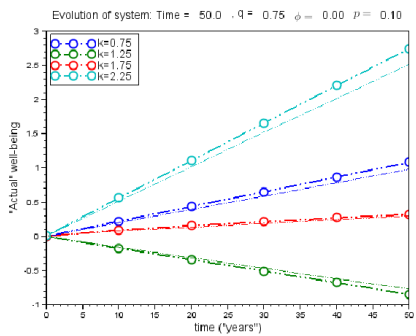

**Figure 2** Evolution of relative well-being for $\phi$=0, for various tradeoff rates with fixed overvaluation factor $q$=0.75. Lines with circles show the relative well-being for $p$=0, while lines without circles correspond to $p$=0.10.

Figure 3 represents the same simulations as in Figure 2, in such a way as to clarify the observed behavior. The diagram at left shows the "drift" of the population over time in the conspicuous/inconspicuous well-being plane. We may think of the dotted lines in the diagram as analogous to "yard lines" on an (American) football field, which indicate the populations' progress as far as anticipated relative well-being. In all cases except $k=1+q$, the circles (which correspond to the population every 10 years, as in Figure 2) are moving "downfield", that is up and to the right, with respect to these dotted "yard lines". This indicates that the average anticipated relative well-being does increase, regardless of $k$. In the cases where $k<1+q$, the conspicuous well-being increases over time at the expense of inconspicuous well-being, while the reverse is true when $k>1+q$. When $k=1+q$, the overvaluation rate exactly matches the tradeoff rate, and no progress in anticipated well-being is possible.

On the other hand, the solid black "yard lines" indicates the population's progress as far as well-being is concerned. Here we see why the fastest progress in relative well-being occurs for the highest tradeoff rate ($k>1+q$). In this case, the agents choose to sacrifice their conspicuous well-being in favor of inconspicuous, since this exchange greatly favors relative well-being. On the other hand, we can also see why the relative well-being decreases for $1<k<1+q$. in this case, although the average anticipated relative well-being is increasing, due to the different alignment of the two sets of "yard lines" the average relative well-being is *decreasing*. This corresponds to the "frog in the pot" syndrome discussed in (Thron 2014), where agents choose *against* their own self-interest because they place too much weight on conspicuous factors. If the tradeoff rate exceeds a threshold ($k>1+q$), then agents will sense the "heat" and stop trying to increase their conspicuous well-being.

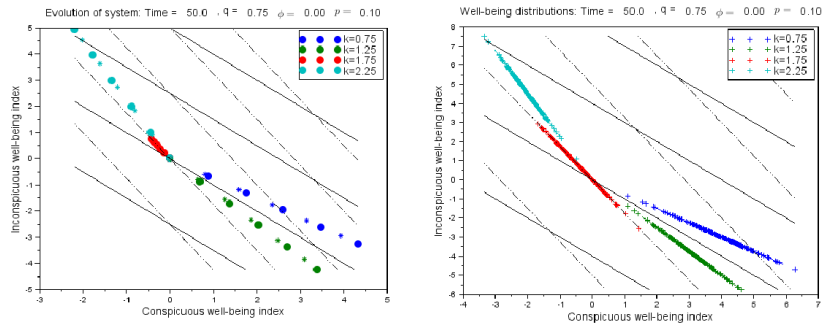

Figure 3 Evolution of relative well-being for $\phi=0$, shown in the conspicuous-inconspicuous well-being plane. In the figure at left, circles show successive "snapshots" of the population's average conspicuous/inconspicuous well-being over time. (Circles in this figure correspond to the circles in Figure 2.) The smaller asterisks in this figure show corresponding snapshots when the turnover probability is set to 0.10. The figure at right shows a scatterplot of the distribution of agents' conspicuous and inconspicuous relative well-beings at time $t=50$. In both figures, the solid black lines indicate lines of constant relative well-being, while the dotted lines indicate lines of constant anticipated relative well-being.

The diagram on the right of Figure 3 shows the distributions of agents' relative well-being at time $t=50$, for different values of $k$. (The distributions are typical for other times as well.) The slopes of the distributions correspond exactly to the corresponding values of $k$, which reflects the fact that $\phi=0$ so that there is a precise linear tradeoff between conspicuous/inconspicuous well-being for new lifestyle choices. By looking at the dotted "yard lines", we can see that when $k<1+q$, the agents with higher conspicuous index values also have higher anticipated well-being. As a result, there is a net pressure on the population towards increasing the conspicuous index. On the other hand, when $k>1+q$ the pressure is in the other direction, towards increasing inconspicuous index. When $k=1+q$, there is no strong pressure either way.

In summary, we have seen that in most cases agents do tend to improve their own actual well-being, *except* in the case where the tradeoff rate exceeds 1 but is dominated by the overvaluation rate ($1<k<1+q$). In this case, agents' imbalanced judgment leads them to increase their conspicuous relative well-being, at the expense of their relative well-being. Figure 3 displays the mechanism which drives these changes.

In the above figures we kept $q$ fixed and varied $k$, and showed that the system behavior depends on the relative sizes of $q$ and $k$. If we fix $k$ and vary $q$, we find that changing $q$ has little effect as long as the size relationship between $k$ and $1+q$ is preserved.

**Behavior with nonzero covariance factor**

When the covariance factor $\phi$ becomes positive the situation changes radically, as shown in Figure 4. (Note the time scale has been expanded compared to Figure 2 to show the changing rates of variation over time.) In this case, during an initial period there is an increase in relative well-being, regardless of tradeoff rate. This initial period can be explained as the result of a sudden diversification of lifestyle choices within the society. Such a situation might arise as a result of sudden economic liberalization, such as occurred in China in the 1980's. However, such "euphoric" periods should be expected to be rare in societies where free economic choice is a given.

Following the initial period of equilibration, for all values of $k>1$ there is a steady decrease in relative well-being. When the turnover probability is positive, large decreases in relative well-being are obtained, especially for larger values of $k$. This is markedly different from the $\phi=0$ case, in which a nonzero turnover probability had only a minor effect.

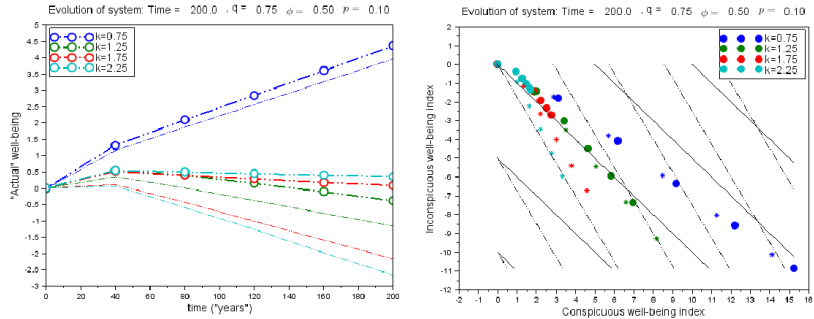

**Figure 4** (*left*) Evolution of relative well-being for fixed $\phi$ (0.5) and $q$ (0.75) for various tradeoff rates. Lines with circles show the relative well-being for $p=0$, while lines without circles correspond to $p=0.10$. (*right*) Evolution of relative well-being for the same scenarios, shown in the conspicuous-inconspicuous well-being plane. Circles show successive "snapshots" of the population's average conspicuous/inconspicuous well-being over time. (The solid circles in this figure correspond to the hollow circles in the figure at right.) The smaller asterisks in this figure show corresponding snapshots when the turnover probability is set to 0.10.

The diagrams in Figure 5 (which is analogous to Figure 3(*right*)) explain the tendencies shown in Figure 4. The covariance ellipses show the regions in the conspicuous-inconspicuous plane where 95% of new lifestyle choices are created, for each scenario. The new choices that correspond to the highest anticipated relative well-being lie near the right upper edges of the covariance ellipses. When agents accept these choices, the mean conspicuous relative well-being index is pushed up, which drives down the mean inconspicuous relative well-being index for new lifestyle opportunities (because of point (4) in the mathematical description of the model above). In all cases, the distribution of agents migrates down and to the right. When $k<1$, the mean relative well-being nonetheless increases because the increase in conspicuous well-being more than offsets the decrease in inconspicuous well-being. But in all other cases, the mean relative well-being decreases, where larger decreases correspond to relatively mild tradeoff rates ($k<1+q$), as in the $\phi=0$ case.

The downward migration is significantly accelerated when a nonzero turnover probability is introduced, as shown in the diagram on the right of Figure 5. When turnovers occur, the only way that agents can recover is by making lifestyle choices with unfavorable tradeoffs, which improve conspicuous well-being index but produce even greater degradation in inconspicuous well-being. Correspondingly, the "frog in the pot" effect occurs for all worse-than-par tradeoff rates ($k>1$), and becomes increasingly severe as $k$ increases.

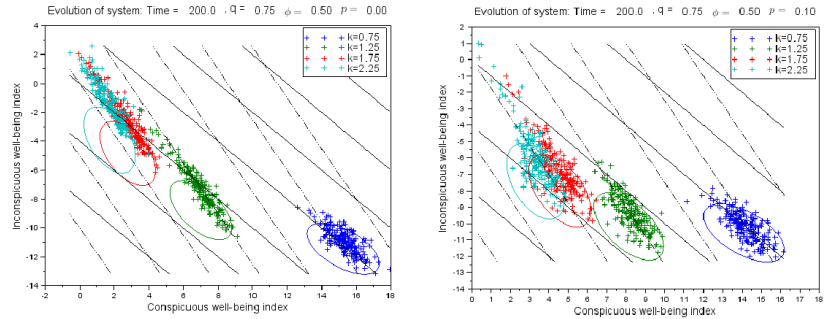

**Figure 5 (*Left*) Scatterplots of the distribution of conspicuous/inconspicuous relative well-being (*C* versus *I*) for various tradeoff rates at time *t*=200, with turnover probability *p*=0. Each ellipse encircles 95% of the new lifestyle choices for the corresponding value of *k* at time *t*=200. (*Right*) Same as the plot at left, except with turnover probability *p*=0.1.**

Figure 6 show the effects of varying overvaluation factors for fixed *k*=1.75. As expected, larger values of *q* lead to larger decreases in well-being. When agents make unbiased lifestyle judgments (*q*=0) the average well-being stabilizes (but if ϕ is decreased to less than 0.35, the average well-being increases, as was seen in Figure 2 and Figure 3). All positive values of *q* cause steady decreases in well-being. Note that the changes in mean *C* and *I* values always follow the *y* = -*kx* line: this is because new lifestyle choices are always offered based on a normal distribution whose mean lies on this line. When turnovers are present, the slopes of the well-being versus time curves decrease significantly, except in the case of very large *q*. When *q* is large, then the presence of turnovers slightly slows the agents' mad dash towards conspicuous prosperity: this is because what agents consider to be "unfavorable" turnovers actually tends to improve their well-being, since their perceptions of well-being are severely unbalanced. In Figure 6, note that all covariance ellipses have the same shape, since all scenarios have the same tradeoff rate.

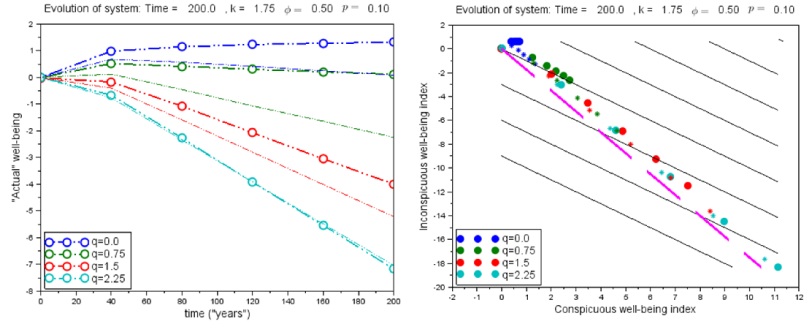

**Figure 6** (*left*) Evolution of relative well-being for fixed $\phi$ (0.5) and $k$ (1.75) for various overvaluation factors. Lines with circles show the relative well-being for $p=0$, while lines without circles correspond to $p=0.10$. (*right*) Evolution of relative well-being for the same scenarios, shown in the *C-I* plane. The dashed magenta line is the lifestyle-choice tradeoff line ($y = -kx$). Circles show successive "snapshots" of the population's average conspicuous/inconspicuous well-being over time. (Solid circles in this figure correspond to the hollow circles in the figure at right.) The smaller asterisks in this figure show corresponding snapshots when the turnover probability is set to 0.10.

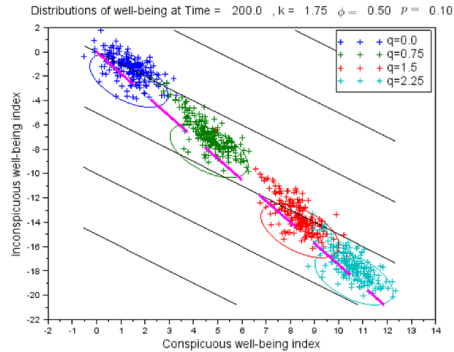

**Figure 7** Scatterplots of the *C* versus *I* distribution for various overvaluation factors at time $t=200$, with $k = 1.75$, $\phi=0.5$, and $p=0.1$. Each ellipse encircles 95% of the new lifestyle choices for the corresponding scenario. The dashed line is the *C-I* tradeoff line, $y=-kx$.

Table 1 summarizes the behavior of the basic model for different model configurations. For each configuration the direction of the trend in actual well-being is indicated by up and down arrows, and multiple arrows are used to indicate the relative sizes of trends. The '+' and '−' entries show the effect on trends when the indicated parameter variation is performed. Three regimes are identified ($k<1$, $1<k<1+q$, and $k>1+q$) which exhibit different qualitative behavior. The table shows the change in behavior in these regimes when nonzero values of uncertainty $\phi$ and turnover $p$ are introduced. We may see how uncertainty and turnover combine to create negative trends in actual well-being whenever overvaluation exists and the conspicuous-inconspicuous tradeoff rate is unfavorable to inconspicuous factors.

**Table 1** Behavior of basic model in different tradeoff / overvaluation regimes (the number of symbols in each entry (↑,↓,+, or −) indicates the relative size of the effect).

|  | $k<1$ | $1<k<1+q$ | $k=1+q$ | $k>1+q$ |
|---|---|---|---|---|
| **$\phi = 0$ case:** | | | | |
| Trend | ↑↑ | ↓↓ | ↑ | ↑↑↑ |
| Change in trend when $p$ increases | − | + | 0 | − |
| Change in trend when $k$ increases | − | + | + | + |
| Change in trend when $q$ increases | 0 | 0 | 0 | 0 |
| **$\phi > 0$, $p=0$ case:** | | | | |
| Trend | ↑↑ | ↓↓ | ↓ | ↓ |
| Change in trend when $p$ increases | − | −− | −−−− | −−− |
| Change in trend when $k$ increases | − | + | + | + |
| Change in trend when $q$ increases | − | − | − | − |
| **$\phi > 0$, $p>0$ case:** | | | | |
| Trend | ↑↑ | ↓↓ | ↓↓↓ | ↓↓↓↓ |
| Change in trend when $p$ increases | − | −− | −−−− | −−− |
| Change in trend when $k$ increases | − | − | − | − |
| Change in trend when $q$ increases | − | − | − | − |

## Variations of the basic model

In this section we discuss two variations to the basic model. First we consider the case where not all agents in the population have the same overvaluation factor; and second, we consider the case where steady upward trends are present in the economic situation.

### Effects of variable overvaluation factor within a population

Until now we have assumed that the overvaluation factor is the same for all agents. We now consider the case when there is a mix of overvaluation factors. In particular, we assign $q=0.5$ to half of the agents, and $q=1.0$ to the other half. The intent of this simulation is to investigate the influence that the two subpopulations have on each other, and to see whether this inhomogeneity has a significant effect on the overall population behavior. Figure 8(*left*) shows that compared to a homogeneous population with $q=0.75$, the mixed population has significantly slower decrease in relative well-being even though the mean $q$ value is the same. In Figure 8(*right*) we compare the distribution of agents at time $t=200$ for three homogeneous scenarios ($q=0.5,1.0$, and $0.75$) with the mixed population. The three covariance ellipses in the figure encircle 95% of the agents for the three given scenarios. The green and red scatter markers correspond to representative random samples of the low-$q$ and high-$q$ subpopulations in the mixed population. We see that many of the low-*q* agents in the mixed population appear to be relatively unaffected by high-*q* agents, since they are located in the same region as most agents in the low-*q* scenario (green ellipse). However, the high-*q* agents within the mixed population seem to be significantly restrained: few of the red scatter points lie in the red ellipse, which represents the location of most agents in an unmixed high-*q* scenario. In the mixed-*q* scenario, the agents are somewhat more spread out than the unmixed scenario with the same average overvaluation factor (blue ellipse.)

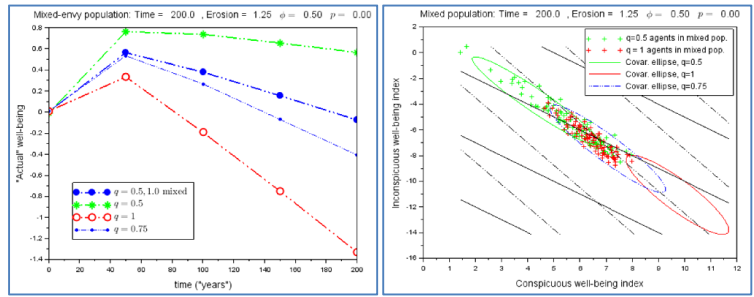

**Figure 8(*left*) Evolution of relative well-being in mixed-*q* population versus high-, low-, and average-*q* unmixed populations, with *p*=0.(*right*) Scatterplots showing the distribution of conspicuous/inconspicuous relative well-being at time T=200.**

Figure 8 shows the case where the turnover probability is zero. When a nonzero turnover probability is introduced, the low-*q* agents in the mixed population are no longer as effective in restraining the rest of the population, as shown in Figure 9. The fact that all agents are subject to the possibility means that low-*q* agents are no longer able to remain in their preferred lifestyle situations, and are pulled down along with the rest of the population. The decrease in relative well-being is only slightly slower than the decrease experienced in an unmixed population with the same average overvaluation factor.

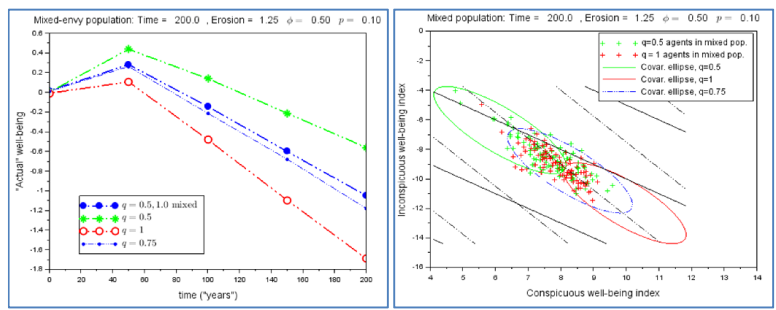

**Figure 9 (*left*) Evolution of relative well-being in mixed-*q* population versus high-, low-, and average-*q* unmixed populations, with *p*=0.1. (*right*) Scatterplots showing the distribution of conspicuous/inconspicuous relative well-being at time T=200.**

**The effect of economic and technological progress**

So far we have not included the possible effects of changes in the agents' situation due to overall economic and technological progress. As a result of such progress, enhancements in conspicuous well-being require fewer resources of time and effort, and are thus less costly in terms of inconspicuous factors. These effects could be reflected in the model through changes over time in the distribution of new lifestyle choices, specified in Point (4) of the mathematical description of the model. Material improvements may serve to modify the tradeoff line for new lifestyle choices. They may reduce the tradeoff rate (for instance, by reducing costs or increasing efficiency); or they may shift the conspicuous-inconspicuous tradeoff line (for instance, through general improvements in healthcare or environmental conditions); or they may cause a combination of both these effects. Accordingly, we reran simulations with the following modification of the new lifestyle choices:

Mean (conspicuous, inconspicuous) indices for new lifestyle choices at time $t$ =
$$(m_C, -(k-ht)(m_C-gt)),$$

where $h>0$ corresponds to the rate of reduction in the tradeoff rate (which is the slope of the conspicuous-inconspicuous tradeoff line), and $g>0$ produces a progressive shift in the conspicuous-inconspicuous tradeoff line (with no change in slope).

Figure 10 shows the effect of growth on the model for $h=0$ and two different values of $g$ ($g=0.01$ and $g=0.02$). It should be mentioned that the numerical value of $g$ should not be interpreted as an economic growth rate, but rather reflects the change in well-being possibilities created by economic and technological enhancement. The figures show that relative well-being still decreases for smaller growth rates, while larger growth rates can overcome the negative trend of the original model. It should not be surprising that well-being always increases rapidly when $k=0.75$, because conspicuous gains are attained at little cost in inconspicuous well-being. In comparing the two graphs in Figure 10, it is especially interesting to note that when $g$ is smaller, introduction of turnovers has a much larger effect  This seems to indicate that slowdowns in economic growth may have a disproportionately adverse effect on average well-being.

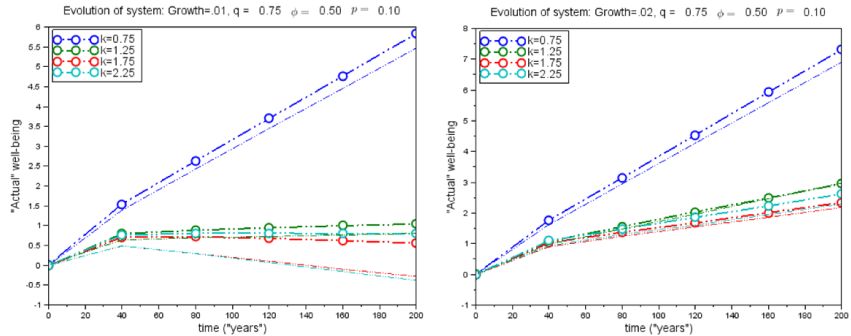

**Figure 10 Evolution of relative well-being for *g*=0.01 (*left*) and *g*=0.02 (*right*) (compare Figure 4, which shows *g*=0 for the same scenarios). Lines with circles show the relative well-being for *p*=0, while lines without circles correspond to *p*=0.10.**

Figure 11 shows the effect of the tradeoff rate decrease parameter *h*, which produces concavity in the well-being trajectories. Sufficiently large positive values of *h* can cause well-being decreases to turn around; and conversely, sufficiently negative values of *h* can reverse increasing trends in well-being.

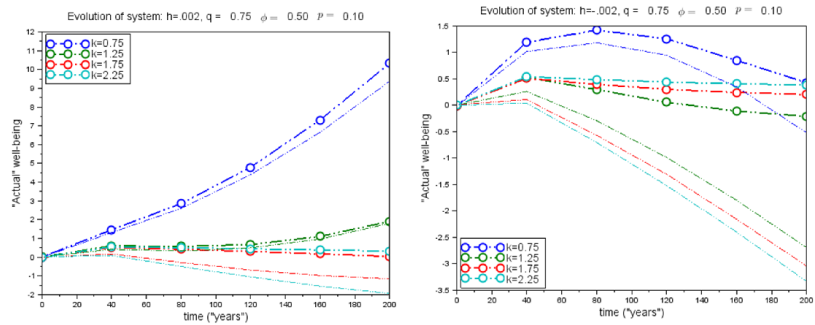

**Figure 11 Evolution of relative well-being for *h*=0.002 (*left*) and *h*=-0.002 (*right*) (compare Figure 4, which shows *h*=0 for the same scenarios). Lines with circles show the relative well-being for *p*=0, while lines without circles correspond to *p*=0.10.**

## Discussion and conclusions

Robert Lane has asserted that there is a natural tendency in free-market societies for well-being to become defined in increasingly materialistic terms (Lane 1999). This is consistent with our simulation results that show conspicuous factors' increasingly dominant contribution to agents' overall well-being. According to our model, the material progress achieved through free-market prosperity may in some circumstances be more than offset by losses in more subtle aspects of quality of life.

American society has experienced several long-term negative social trends that have continued despite their obvious detrimental effects. A famous example is the drift away from close-knit community in the United States, as documented by Robert Putnam (Putnam 2001). Other examples include trends towards increasing percentage of single households (Vespa et. al. 2013), family instability (Furstenburg 1990), obesity (Flegal et. al. 1998), crime (Smith 1995), alcohol abuse (Grant et. al. 2004) and reduced sleep duration for full-time workers (Knutson et. al. 2010). Similar negative trends in social indicators such as divorce, crime, and delinquency rates occurred in China during its period of exceptional economic growth (Wang 2006; Wang & Zhou 2010). Our model suggests that such trends may be at least partly the result of the neglect of inconspicuous psychological and social contributions to well-being, in favor of conspicuous prosperity.

According to the model, the basic mechanism that produces decreases in a population's well-being is due to unbalanced judgment (overvaluation) on the part of agents that make lifestyle choices; coupled with the fact that those choices involve tradeoffs between conspicuous and inconspicuous well-being.

This basic mechanism is exacerbated by a number of factors. Markets that create lifestyle possibilities respond directly to conspicuous rather than inconspicuous factors. As a result, continuing decreases in overall well-being occur when conspicuous gains are earned at the expense of even greater losses in inconspicuous factors. If economic or personal misfortunes occur, then agents are not able to recover their previous level of well-being, due to market shifts that favor conspicuous at the expense of inconspicuous. This seems to indicate that seemingly minor levels of misfortune may cause progressive decreases in a population's average well-being, even when agents do not overvalue conspicuous aspects of well-being.

The natural variations in lifestyle choices available at any given time appears to play a part in promoting decreases in well-being, which as far as the author knows has not previously been recognized in the literature. Such variations can obscure net unfavorable tradeoffs between conspicuous and inconspicuous factors. Opportunistic individuals may obtain short-term gains in well-being, but in the long term the entire population suffers as others try to duplicate their success. This process could be compared to a lottery which individuals continue to play despite the certainty of net loss for the population as a whole.

The presence of a cognizant subpopulation that does not overvalue conspicuous well-being can slow, but not stop, the decay of well-being due to the unbalanced judgments of the rest of the population. In the end, the cognizant subgroup's well-being is pulled down along with the rest of the population, especially when the risk of turnovers is included in the model. Turnovers serve to homogenize the population, because individuals trying to recover their level of well-being after a personal stumble are forced to choose from the current distribution of available options.

Economic and technological progress can offset the negative effects that we have discussed above. If this progress is sufficient, then the population's average well-being is not permanently affected by turnovers that may be encountered by individuals within the population. However, once progress dips below a critical level, then disproportionate decreases in well-being are to be expected,

Our results suggest that warning signs that indicate the possibility of decreasing well-being are: (1) individuals within the society seriously overvalue conspicuous aspects of well-being in their lifestyle choices; (2) tradeoffs between conspicuous and inconspicuous well-being are sufficiently moderate that they are masked by agents' overvaluation of conspicuous prosperity; (3) job and consumer markets are insensitive to inconspicuous well-being; (4) a non-negligible proportion of agents within the population experience economic or personal setbacks during each fixed time period.

The model has significant implications as far as social policy. It suggests a comprehensive strategy that addresses several of the four conditions listed in the previous paragraph. Possible actions include the use of political, religious, and popular cultural media platforms to promote (and even overemphasize) appreciation of inconspicuous aspects of well-being. Schools could lay greater educational emphasis on the arts, literature, and other "nonproductive" but personally enriching endeavors. Other possible approaches include taxation or regulatory legislation that discourages excessive consumption, overwork, and nomadic job-switching behavior among career-seeking professionals.